\documentclass[journal]{IEEEtran}

\usepackage[T1]{fontenc}
\usepackage{xspace}
\usepackage{cite}

\usepackage[cmintegrals]{newtxmath}

\newcommand{\inputtikz}[1] {
  \includegraphics{#1.pdf}
}

\newcommand{\E}{\ensuremath{\mathbb{E}}}

\newcommand{\Rlossn}{\ensuremath{R_{\text{loss},n}}\xspace}
\newcommand{\AIRn}{\ensuremath{\text{AIR}_n}\xspace}

\usepackage[utf8]{inputenc}
\usepackage{amsmath}

\usepackage{xcolor}

\newlength{\MyFigureWidth}
\newlength{\MyFigureHeight}
\usepackage{mathtools}
\DeclarePairedDelimiter\abs{\lvert}{\rvert}%
\let\norm\relax
\DeclarePairedDelimiter\norm{\lVert}{\rVert}%
\let\floor\relax
\DeclarePairedDelimiter\floor{\lfloor}{\rfloor}
\let\ceil\relax
\DeclarePairedDelimiter\ceil{\lceil}{\rceil}
\makeatletter
\let\oldabs\abs
\def\abs{\@ifstar{\oldabs}{\oldabs*}}
\let\oldnorm\norm
\def\norm{\@ifstar{\oldnorm}{\oldnorm*}}
\let\oldfloor\floor
\def\floor{\@ifstar{\oldfloor}{\oldfloor*}}
\let\oldceil\ceil
\def\ceil{\@ifstar{\oldceil}{\oldceil*}}
\makeatother

\usepackage[hidelinks]{hyperref} 

\begin{document}
\title{Fiber Nonlinearity Mitigation by Short-Length\\Probabilistic Constellation Shaping\\for Pilot-Aided Signaling}

\author{Tobias Fehenberger, Helmut Griesser, and Jörg-Peter Elbers%
\thanks{The authors are with ADVA, Munich, Germany.\\E-mails: tfehenberger@adva.com, hgriesser@adva.com, jelbers@adva.com.}
}

{}

\maketitle

\begin{abstract}
Probabilistic constellation shaping (PCS) offers a significant performance improvement over uniform signaling. It was recently discovered that long blocks are not required to achieve maximum shaping gain when transmitting over the nonlinear fiber channel because short-length PCS effectively mitigates fiber nonlinear interference (NLI). The reason for this behavior is that short-length PCS implicitly induces some temporal properties in the shaped transmit sequence that are beneficial for the fiber-optic channel. To achieve robust data-aided digital signal processing of high-order QAM, periodic quaternary phase shift keying pilots are typically inserted into the high-order QAM transmit sequence. In this work, we investigate in simulations the effect of such pilot-aided signaling on NLI mitigation. Albeit modifying the temporal properties of the shaped transmit sequence, a pilot rate of 1/32 is found to not alter the beneficial effects of short-length PCS. The operation meaning of this finding is that even with pilot-aided signaling, long PCS block lengths are not required for maximum shaping gain.
\end{abstract}

\begin{IEEEkeywords}
Probabilistic Constellation Shaping, Fiber Nonlinearities, Coded Modulation
\end{IEEEkeywords}

\IEEEpeerreviewmaketitle

\section{Introduction}
Probabilistic constellation shaping (PCS) is a powerful technique to closely approach the Shannon limit by reducing the performance gap of uniformly distributed quadrature amplitude modulation (QAM). Another PCS feature that is at least as appealing as the shaping gain is rate adaptivity. With fixed QAM order and forward error correction (FEC) overhead, the net throughput can be varied by modifying the rate of the shaping device used in PCS. These two features make PCS a digital signal processing (DSP)building block that has made, or will make, its way into many commercial high-performance  modules.

Many modern PCS implementations are based on the probabilistic amplitude shaping (PAS) framework \cite{Boecherer2015TransComm_ProbShaping}. A key device of PAS is a distribution matcher (DM), such as a constant-composition distribution matcher (CCD) \cite{Schulte2016TransIT_DistributionMatcher}, that performs a fixed-length block-wise mapping from uniform data bits to shaped amplitudes. For latency and complexity reasons, short DM blocks are beneficial, yet this typically brings about a significant performance penalty due to increased DM rate loss at small block lengths. Hence, advanced DM methods with reduced rate loss, such as multiset-partition DM \cite{Fehenberger2019TCOM_MPDM }or enumerative sphere shaping\cite{GultekinEntropy2020_ShaperComparison}, have been devised. Many of these schemes were  studied and optimized for a linear channel with additive noise only, which obviously neglects the impact of fiber nonlinear interference (NLI).

It was recently demonstrated \cite{amariIntroducingEnumerativeSphere2019,goossens2019first,civelli2020interplay,Fehenberger2020OFC_ShapingNLI} that PCS signaling can have a positive impact on the NLI magnitude because short-length PCS can, under certain conditions and against previous belief \cite{Fehenberger2016JLT_ShapingQAM}, lead to NLI mitigation. This means that there exists an optimal shaping block length where short-length NLI reduction and long-length rate loss improvements are balanced out. This effect is attributed to temporal correlations in the transmit sequence that are implicitly introduced by short-length shaping. For constant-composition sequences, a detailed analysis can be found in \cite{Fehenberger2020JLT_CC}.

In our previous work \cite{Fehenberger2020OFC_ShapingNLI}, we have shown that nonlinearity mitigation due to short-length PCS vanishes when the transmit sequences are interleaved as this shuffling removes the beneficial temporal structure introduced by short DMs. In this work, we investigate the case when quaternary phase shift keying (QPSK) pilots are periodically inserted in the shaped transmit sequence, which is typically done to enable robust DSP for high-order QAM. Simulations of wavelength division multiplexing (WDM) transmission over a multi-span fiber system show that inserting QPSK pilots at a rate of 1/32, which is compatible to the 400ZR implementation agreement, does not affect the NLI mitigation, despite the temporal structure of the transmit sequence being altered. The operational implication of the presented study is that there is no need to spend resources for long-block-length PCS implementations for nonlinear fiber transmission even when QPSK pilots are inserted.

\begin{figure*}
\begin{center}
\inputtikz{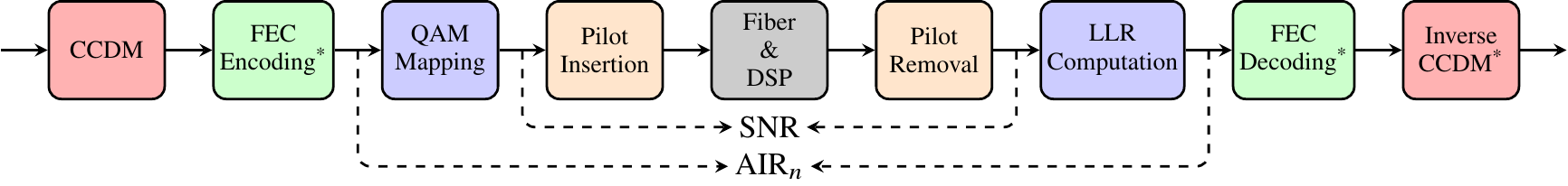}
\end{center}
\vspace*{-\baselineskip}
\caption{Simplified block diagram of the considered optical transmission system. The blocks marked with an asterisk are omitted in the simulations or, in the case of FEC encoding, emulated.}
\label{fig:block_diagram}
\end{figure*}

\begin{figure*}
  \centering
  \inputtikz{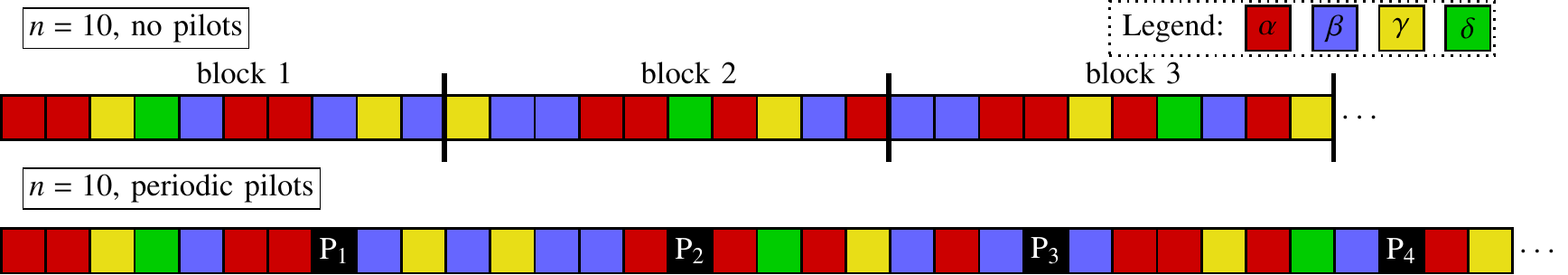}
  \caption{\label{fig:CCDM_block_example} Illustration of three CCDM blocks of length $n=10$. Top: without pilots; bottom: with periodically inserted pilots.}
\end{figure*}

\section{Simulation Setup}
The effects of short-length PS on the SNR and on bit-wise AIRs after fiber transmission and DSP are studied for different CCDM block lengths. A block diagram of the communication system is shown in Fig.~\ref{fig:block_diagram}. We consider the case with no pilots symbols and also the more practically more relevant scenario where QPSK pilots are periodically inserted into the transmit sequence. The pilots are inserted at a rate of 1/32, i.e., a block of 31 64-QAM symbols is followed by a single QPSK pilot, which is compatible to the 400ZR implementation agreement \cite{400ZR}. Note that we do not consider a long header sequence of several a-priori known symbols at the beginning of a block as we focus our analysis on the impact of pilot that break up the temporal CCDM properties. Also, FEC is not included in the simulations as it does not affect the presented analysis. Instead of FEC encoding, the sign bits for PAS are generated randomly with a uniform distribution.

We simulated 324000 64QAM-symbols per polarization. The total simulation sequence comprises several CCDM blocks whose length is varied from $n=10$ up to $n=400$ amplitude symbols. The shaped amplitude distribution of 64QAM is fixed to $[0.4, 0.3, 0.2, 0.1]$. This is by no means the optimal distribution for the considered setup, but this allows to focus on nonlinear fiber effects. With this setup, all QAM transmit sequences have the same average distribution---the only difference is the length of the CCDM blocks that constitute that overall sequence and whether or not pilots are included. A schematic of the transmit sequence is illustrated in Fig.~\ref{fig:CCDM_block_example} for CCDM with $n=10$ and 64QAM, which corresponds to four shaped amplitudes denotes as $\alpha$, $\beta$, $\gamma$, and $\delta$. The top part of Fig.~\ref{fig:CCDM_block_example} shows three blocks of this length, where due to the constant-composition principle, $\alpha$ must occur exactly four times in each block, $\beta$ three times etc. Note that this introduces temporal properties in the overall compound sequence that would not be present for longer $n$. For instance, using a small $n$ limits the length of identical-symbol runs, or, equivalently, induces some inherent shuffling that has been found to be beneficial for reducing fiber NLI \cite{Fehenberger2020SPPCom_ShapingNLI,Fehenberger2020JLT_CC,Fehenberger2020OFC_ShapingNLI}. The bottom part of Fig.~\ref{fig:CCDM_block_example} shows the same CCDM sequences but with pilots inserted (for illustration purposes with a pilot rate smaller than 1/32).

To investigate whether these QPDK pilots alter the NLI mitigating effect, an idealized polarization-multiplexed  WDM fiber system is simulated for optical transmission. The shaped sequences in both polarizations are generated from independent random data, and each WDM channel is modulated individually. Seven WDM channels at 42 GBd symbol rate on a 50 GHz grid are simulated with an optimal per-channel transmit power of 1~dBm. After digital root-raised cosine pulse shaping with 10\% roll-off, the dual-polarization WDM signal is transmitted over 10 spans of 80~km standard single-mode fiber ($\alpha=0.2$dB/km, $\gamma=1.37$~1/W/km, $D=17$~ps/nm/km). The span loss of 16~dB is compensated by an Erbium-doped fiber amplifier with 6 dB noise figure. Signal propagation over the fiber is simulated with the split-step Fourier method using a 100~m step size. At the receiver, chromatic dispersion is compensated, the center channel is ideally filtered using a matched filter, and the nonlinearity-induced constant phase rotation is ideally compensated.

We consider two different figures of merit, which are effective SNR and achievable information rate (AIR), both computed from the 64QAM symbols only (see also \ref{fig:block_diagram}). The effective SNR, averaged over both polarizations, is estimated from the unit-energy transmitted data $x$ and the received symbols $y$ as $1/\text{var}(y-x)$ where $\text{var}(\cdot)$ denotes variance. As SNR does not reflect the finite-length DM rate loss or in general the throughput improvement from PCS, an AIR for bit-metric decoding (BMD) and for a finite-length DM of length $n$ is computed as \cite[Appendix]{Fehenberger2019TCOM_MPDM}
\begin{equation}\label{eq:air}
\AIRn = \left[ H(\mathbf{C}) - \sum_{i}^{m} H(C_i|Y) \right] - \Rlossn.
\end{equation}
The first part in square brackets is the BMD rate with $\mathbf{C}=(C_1,\dots,C_m)$ representing the $m$ coded bit levels of the considered 64QAM format, $H(\cdot)$ denoting entropy and the channel output being $Y$. The rate loss \Rlossn is defined as the amplitude entropy minus the CCDM rate, i.e.,
\begin{equation}
\Rlossn = H(A) - \frac{k}{n},
\end{equation}
where $k$ is the number of CCDM input bits. In general, \Rlossn{} decreases with distance. We note that \AIRn of \eqref{eq:air} is achievable with a finite-length DM that has rate loss \Rlossn and with capacity-achieving FEC. If an actual FEC is implemented, the required SNR for successful decoding will be decreased by approximately the FEC coding gap \cite{Fehenberger2019TCOM_MPDM}.

\section{Numerical Results}
\begin{figure}[t]
\begin{center}
\inputtikz{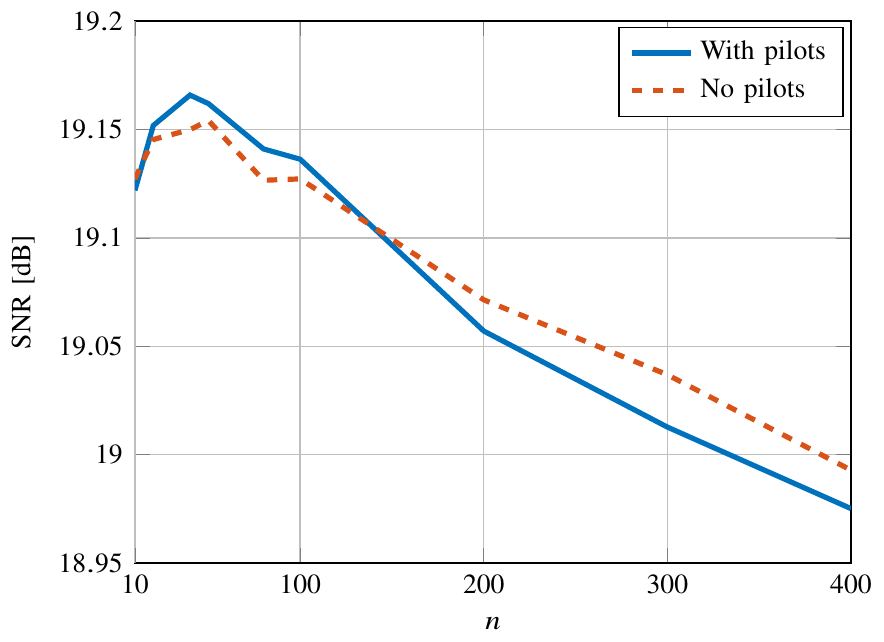}
\end{center}
\vspace*{-\baselineskip}
\caption{SNR in dB versus CCDM block length $n$ for 64QAM signaling with and without pilots periodically inserted pilots at rate of 1/32. The SNR estimation is performed using only the 64QAM symbols.}
\label{fig:SNR_vs_n}
\end{figure}

Figure~\ref{fig:SNR_vs_n} shows the SNR estimated from the shaped 64QAM symbols after DSP versus a range of CCDM block lengths from $n=50$ to $n=400$. For very small block lengths less than 50 symbols---which are practically irrelevant for CCDM due to the high rate loss---a marginal SNR increase is observed, which we mostly attribute to numerical fluctuations. For larger $n$, we observe that the SNR is continuously decreasing and differs by almost 0.2~dB over the depicted block length range. In previous work \cite{Fehenberger2020OFC_ShapingNLI}, up to 0.8~dB SNR difference have been observed when extending the maximum block length to $n=5000$ symbols. The reason for this behavior is, as already mentioned above, that short-length CCDM imposes some restrictions on the properties of the compound sequence that manifest as inherent shuffling of the amplitudes and a limited length of identical-symbol runs\cite{Fehenberger2020JLT_CC}. Note that the SNR dependence on the block length can only be observed for a nonlinear channel, but not for a linear channel with, for example, additive white Gaussian noise where the estimated SNR is independent of $n$. It becomes obvious from Fig.~\ref{fig:SNR_vs_n} that the inverse proportionality of SNR with $n$ is also present when pilots are periodically inserted. 

\begin{figure}[t]
\begin{center}
\inputtikz{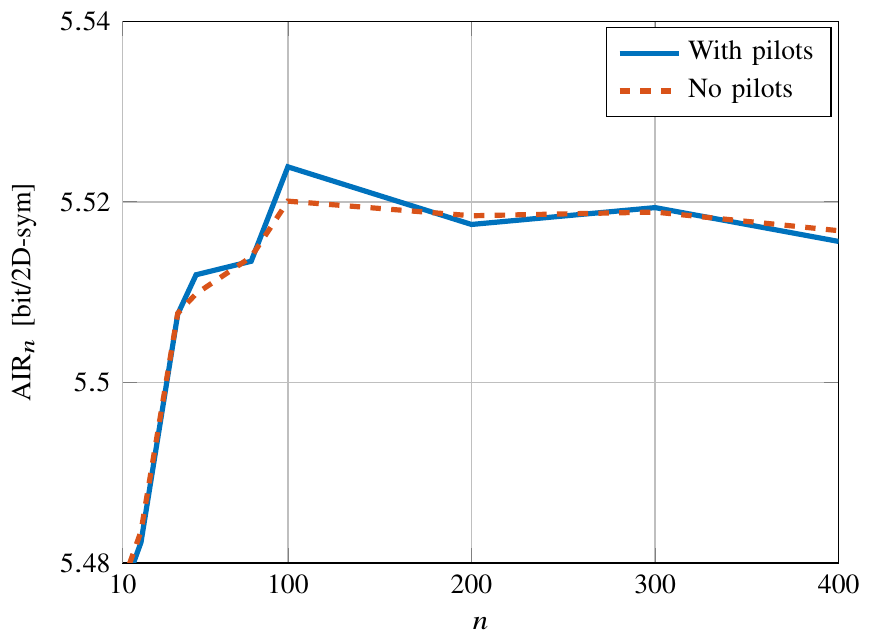}
\end{center}
\vspace*{-\baselineskip}
\caption{\AIRn in bits per 2D-symbol versus CCDM block length $n$, again with only the 64QAM payload used for estimation of \AIRn.}
\label{fig:GMI_vs_n}
\end{figure}

So far we have evaluated only SNR. The much more relevant figure of merit of a digital communication system, however, is the AIR describing the maximum attainable throughput for capacity-achieving FEC. In Fig.~\ref{fig:GMI_vs_n}, we show \AIRn{} defined in \eqref{eq:air}, which assumes ideal binary FEC and also captures the finite-length CCDM rate loss. We observe that the GMI approximately saturates at block lengths beyond $n=100$. The reason for this is that the SNR loss and the rate loss improvement that both occur  for increasing $n$ effectively cancel out. The practical implication of this result is that using very long block lengths might not be required for maximizing performance of a fiber-optic communication system with PCs. This also means that the extra DSP complexity that is put into achieving extremely long block lengths---which would give the best performance for a linear channel but not for the nonlinear fiber channel---could be saved.

\section{Conclusions}
We have analyzed the block-length dependence of SNR and AIR in numerical simulations of a fiber-optic communication systems with PCS. In addition to previous work, we have included in our analysis a periodic insertion of QPSK pilots, which is typically done for robust data-aided DSP of high-order QAM. The NLI mitigation due to short-length is found to remain largely unchanged despite the QPSK pilots breaking up the temporal properties of the constant-composition sequences. The conclusion is that for nonlinear fiber transmission, the additionally required complexity to realize long-block-length PCS can be saved.

\bibliographystyle{IEEEtran}

\begin{thebibliography}{10}
\providecommand{\url}[1]{#1}
\csname url@samestyle\endcsname
\providecommand{\newblock}{\relax}
\providecommand{\bibinfo}[2]{#2}
\providecommand{\BIBentrySTDinterwordspacing}{\spaceskip=0pt\relax}
\providecommand{\BIBentryALTinterwordstretchfactor}{4}
\providecommand{\BIBentryALTinterwordspacing}{\spaceskip=\fontdimen2\font plus
\BIBentryALTinterwordstretchfactor\fontdimen3\font minus
  \fontdimen4\font\relax}
\providecommand{\BIBforeignlanguage}[2]{{%
\expandafter\ifx\csname l@#1\endcsname\relax
\typeout{** WARNING: IEEEtran.bst: No hyphenation pattern has been}%
\typeout{** loaded for the language `#1'. Using the pattern for}%
\typeout{** the default language instead.}%
\else
\language=\csname l@#1\endcsname
\fi
#2}}
\providecommand{\BIBdecl}{\relax}
\BIBdecl

\bibitem{Boecherer2015TransComm_ProbShaping}
G.~B{\"{o}}cherer, P.~Schulte, and F.~Steiner, ``{Bandwidth efficient and
  rate-matched low-density parity-check coded modulation},'' \emph{{IEEE}
  Transactions on Communications}, vol.~63, no.~12, pp. 4651--4665, Dec. 2015.

\bibitem{Schulte2016TransIT_DistributionMatcher}
P.~Schulte and G.~B{\"{o}}cherer, ``{Constant composition distribution
  matching},'' \emph{{IEEE} Transactions on Information Theory}, vol.~62,
  no.~1, pp. 430--434, Jan. 2016.

\bibitem{Fehenberger2019TCOM_MPDM}
T.~Fehenberger, D.~S. Millar, T.~Koike-Akino, K.~Kojima, and K.~Parsons,
  ``{Multiset-partition distribution matching},'' \emph{{IEEE} Transactions on
  Communications}, vol.~67, no.~3, pp. 1885--1893, Mar. 2019.

\bibitem{GultekinEntropy2020_ShaperComparison}
Y.~C. Gültekin, T.~Fehenberger, A.~Alvarado, and F.~M.~J. Willems,
  ``{Probabilistic shaping for finite blocklengths: distribution matching and
  sphere shaping},'' \emph{Entropy}, vol.~22, no. 581, pp. 1--31, May 2020.

\bibitem{amariIntroducingEnumerativeSphere2019}
A.~Amari, S.~Goossens, Y.~C. Gultekin, O.~Vassilieva, I.~Kim, T.~Ikeuchi,
  C.~Okonkwo, F.~M.~J. Willems, and A.~Alvarado, ``{Introducing enumerative
  sphere shaping for optical communication systems with short blocklengths},''
  \emph{arXiv:1904.06601 [cs, math]}, Apr. 2019.

\bibitem{goossens2019first}
S.~Goossens, S.~Van~der Heide, M.~Van~den Hout, A.~Amari, Y.~C. G{\"u}ltekin,
  O.~Vassilieva, I.~Kim, T.~Ikeuchi, F.~M. Willems, A.~Alvarado \emph{et~al.},
  ``First experimental demonstration of probabilistic enumerative sphere
  shaping in optical fiber communications,'' in \emph{OptoElectronics and
  Communications Conference (OECC)}, 2019.

\bibitem{civelli2020interplay}
S.~Civelli, E.~Forestieri, and M.~Secondini, ``Interplay of probabilistic
  shaping and carrier phase recovery for nonlinearity mitigation,'' \emph{arXiv
  preprint arXiv:2009.01135}, 2020.

\bibitem{Fehenberger2020OFC_ShapingNLI}
T.~Fehenberger, H.~Griesser, and J.-P. Elbers, ``{Mitigating fiber
  nonlinearities by short-length probabilistic shaping},'' in \emph{Proc.
  Optical Fiber Communication Conference \mbox{(OFC)}}, San Diego, CA, USA,
  Mar. 2020.

\bibitem{Fehenberger2016JLT_ShapingQAM}
T.~Fehenberger, A.~Alvarado, G.~B{\"o}cherer, and N.~Hanik, ``{On probabilistic
  shaping of quadrature amplitude modulation for the nonlinear fiber
  channel},'' \emph{{IEEE/OSA} Journal of Lightwave Technology}, vol.~34,
  no.~22, pp. 5063--5073, Nov. 2016.

\bibitem{Fehenberger2020JLT_CC}
T.~{Fehenberger}, D.~S. {Millar}, T.~{Koike-Akino}, K.~{Kojima}, K.~{Parsons},
  and H.~{Griesser}, ``{Analysis of nonlinear fiber interactions for
  finite-length constant-composition sequences},'' \emph{{IEEE/OSA} Journal of
  Lightwave Technology}, vol.~28, no.~3, pp. 457--465, Jan. 2020.

\bibitem{400ZR}
I.~Lyubomirsky, ``{400GBASE-ZR PCS/PMA baseline proposal},'' in \emph{{IEEE
  P802.3cn Task Force Meeting}}, 2019.

\bibitem{Fehenberger2020SPPCom_ShapingNLI}
T.~Fehenberger, ``{On the impact of finite-length probabilistic shaping on
  fiber nonlinear interference},'' in \emph{Signal Processing in Photonic
  Communications (SPPCom)}, Montreal, Canada, Jul. 2020.

\end{thebibliography}

\end{document}